\documentclass[10pt]{article}  
\usepackage{amssymb,amsmath}  
\usepackage{url}
\usepackage[colorlinks=true, pdfstartview=FitV, linkcolor=blue, citecolor=blue, urlcolor=blue]{hyperref}
\usepackage{upgreek}

\def\div{\nabla\cdot } 
\def\pl{\partial}

\def\={\equiv}

\newcommand{\dc}{_{\scriptscriptstyle\infty}}

\newcommand{\blue}{\color{blue}}

\newcommand{\rt}{(\3r,t)}

\newcommand{\rrm}[1]{{\rm #1}}

\newtheorem{rem}{Remark}

\newcommand{\bib}{\bibitem}

\newcommand{\nt}{\notag}
\newcommand{\ci}{\cite}
\newcommand{\lab}{\label}

\newcommand{\eq}{\eqref}

\newcommand{\lp}{\left(}
\newcommand{\rp}{ \right)}

\newcommand{\la}{\langle\,}

\newcommand{\ra}{\,\rangle}

\newcommand{\harr}[1]{\smash{\mathop{\hbox to .5in{\ \rightarrowfill\ }}
      \limits^{#1}}}

\newcommand{\0}[1]{{(#1)}}

\newcommand{\3}[1]{{\boldsymbol #1}}
\newcommand{\bb}[1]{{\boldsymbol{\bar #1}}}
\newcommand{\bh}[1]{{\boldsymbol{\hat #1}}}

\newcommand{\fr}[1]{{\mathfrak #1}}

\newcommand{\6}[1]{_{\scriptscriptstyle#1}}
\newcommand{\7}[1]{{\bar#1}}

\newcommand{\9}[1]{^{\scriptscriptstyle#1}}

\newcommand{\mb}[1]{{\mathbf #1}}
\newcommand{\bmb}[1]{{\mathbf{\bar #1}}}

\def\a{\alpha}

\def\d{\delta} 
\def\e{\varepsilon}

\def\vf{\varphi}

\def\m{\mu} 

\def\o{\omega} 
\def\p{\pi}

\def\r{\rho}
 
\def\s{{\sigma}} 

\def\t{\tau}

\def\D{\Delta}

\newcommand{\db}{{\,{\rm d}\kern-1.6ex-}}
\newcommand{\dir}{{\pl\kern-1.2ex {/}}}
\newcommand{\dd}{{\rm d}}

\newcommand{\curl}{\nabla\times}

\newcommand{\ie}{{\it i.e., }}

\def\iff{\ \Leftrightarrow\ }
  
\newcommand{\imp}{\ \Rightarrow\ }
\newcommand{\inv}{^{-1}}

\newcommand{\ir}{\int_{-\infty}^\infty}  

\newcommand{\lra}{\leftrightarrow}

\newcommand{\plra}{\pl^{\kern-1.25ex^\lra}}
\newcommand{\qq}{\quad}

\newcommand{\sgn}{{\,\rm Sgn \,}}
 
\newcommand{\sr}{\sqrt}

\newcommand{\orr}{{(\3r)}}

\def\Xint#1{\mathchoice
   {\XXint\displaystyle\textstyle{#1}}%
   {\XXint\textstyle\scriptstyle{#1}}%
   {\XXint\scriptstyle\scriptscriptstyle{#1}}%
   {\XXint\scriptscriptstyle\scriptscriptstyle{#1}}%
   \!\int}
\def\XXint#1#2#3{{\setbox0=\hbox{$#1{#2#3}{\int}$}
     \vcenter{\hbox{$#2#3$}}\kern-.5\wd0}}

\def\ppint{\Xint-}

\def\VE{\vfill\eject}

\def\ref#1{(#1)}
\def\dc{\6\infty}

\begin{document}

\title{Completing the complex Poynting theorem: \\
Conservation of reactive energy in reactive time
}

\author{{Gerald Kaiser\,\thanks{Supported by AFOSR Grant \#FA9550-12-1-012.}}\\
\href{http://wavelets.com}{Center for Signals and Waves}\\ Portland, OR\\
kaiser@wavelets.com
}

\maketitle

\begin{abstract}\noindent
The complex Poynting theorem is extended canonically to a \it time-scale domain \rm $(t, s)$ by replacing the \it phasors \rm $e^{i\o t}\mb X\orr$ of time-harmonic fields $\3X$ ($\3E,\3H,\3J,\cdots$) with the \it analytic signals \rm $\mb X(\3r, t+is)$ of  fields $\3X\rt$ with general time dependence.  The parameter $s>0$ is shown to play the role of a \it time resolution scale, \rm and the extended Poynting theorem splits into two conservation laws: its real part gives the \it conservation in $t$ \rm of the \it scale-averaged \rm  (over $\D t\sim \pm s$) \it active energy \rm at fixed $s$, and its imaginary part gives the \it conservation in $s$ \rm of the scale-averaged \it reactive energy \rm at fixed $t$. This motivates the interpretation of $s$ as \it reactive time, \rm  measured in \it seconds reactive \rm [sr], giving a rational basis for the \it ampere reactive \rm [Ar] and \it volt-ampere reactive \rm [VAr] units used to measure reactive current and reactive power: [Ar]=[C/sr] and [ VAr] = [J/sr]. At coarse scales (large $s$, slow time), where the system reduces to the \it circuit level, \rm this may have applications to the theory of electric power transmission and conditioning.  At fine scales (small $s$, fast time) it describes reactive energy dynamics in radiating systems.
\end{abstract}

\VE

\tableofcontents

\section{Introduction}\label{S:intro}

Although reactive electromagnetic field energy is often discussed in the physics and engineering literature, there seems to be no actual \it expression \rm for it on par with that for ordinary field energy, which will here be called \it active energy \rm for distinction. There are only expressions for indirectly related variables, such as the reactance of a circuit,  computed from the complex Poynting theorem (see \eq{CP}), hence valid only for time-harmonic fields. For example, Jackson \ci[page 265]{J99} writes: ``It is a complex equation whose real part gives the conservation of energy for the time-averaged quantities and whose imaginary part \it relates to \rm  the reactive or stored energy and its alternating flow.''

I propose here a unified theory of active and reactive energy flow for electromagnetic fields with general time dependence, based solely on Maxwell's equations.

Given an electromagnetic field $\3E, \3H$, the key is to continue its positive-frequency parts $\mb E,\mb H$ analytically in time to $\t=t+is, s>0$. This gives a time-domain version of the \it phasor representation \rm with an additional parameter $s$, which is shown to have a natural interpretation as a \it time resolution scale \rm with $\D t\sim s$.  Next, a generalization of the complex Poynting theorem to the time-scale domain is derived. Its real part gives conservation in the \it real time \rm $t$ of the scale-averaged active energy, and its imaginary part gives conservation in the \it imaginary time \rm $s$ of the scale-averaged reactive energy. It is proposed to interpret $s$ as \it reactive time, \rm tracking the lags and leads associated with reactive energy, to be measured in \it seconds reactive \rm  [sr]. This explains the \it ampere reactive \rm [Ar] and  \it volt-ampere reactive \rm [VAr] units, used to measure reactive current and reactive power, as \it coulombs per second reactive \rm [C/sr] and \it Joules per second reactive \rm [J/sr], respectively. From this point of view, active and reactive energy are on equal footing (both measured in Joules) while active and reactive 
power and energy flux are distinguished only by their respective time variables. Some potential applications are discussed in the final section.

\section{Vector fields in the analytic time-scale domain}

Let $\3X\rt$ be a space-time vector field with temporal Fourier representation
\begin{align*}
\3X\rt=\frac1{2\p}\ir\dd\o\, e^{i\o t}\3X_\o\orr, && \3X_\o\orr=\ir\dd t\,e^{-i\o t}\3X\rt.
\end{align*}
$\3X\rt$ and $\3X_\o\orr$ are called the \it time-domain \rm and \it frequency-domain \rm representations of the field, respectively.  We allow $\3X$ to include a \it static \rm field $\3X\dc\orr$ (DC component), so that
\begin{align}\lab{XX'}
\3X\rt=\3X\dc\orr+\3X'\rt,
\end{align}
with Fourier transform
\begin{align}\lab{XX'o}
\3X_\o\orr=2\p\d\0\o\3X\dc\orr+\3X_\o'\orr,\ \text{where}\  \3X_0'\orr=\30.
\end{align}
We assume throughout that $\3X$ is real, hence $\3X\dc$ is real and $\3X_\o'$ satisfies the reality condition
\begin{align}\lab{rc}
\bb X_\o'\orr=\3X_{-\o}'\orr
\end{align}
where the bar denotes complex conjugation. Thus it is unnecessary to specify the negative-frequency components of $\3X'$ independently. This motivates the following construction, known as the \it analytic signal \rm of $\3X'$. Let
\begin{align}\lab{X0}
\mb X'(\3r,t)\=\frac1\p \int_0^\infty\dd\o\,e^{i\o t}\3X_\o'\orr
\end{align}
be (twice) the \it positive-frequency part \rm of $\3X'$ and
\begin{align}\lab{X00}
\mb X\rt=\3X\dc\orr+\mb X'\rt.
\end{align}
The frequency-domain representation of $\mb X$ is 
\begin{align}\lab{Xo}
\mb X_\o\orr&=2\p\d\0\o\3X\dc\orr+2H\0\o\3X_\o'\orr,
\end{align}
where $H\0\o$ is the Heaviside step function.

\rem\rm\label{R:phasors} Observe the notational distinction between the original fields $\3X',\3X_\o'$ and their analytic signals $\mb X', \mb X_\o'$. For a time-harmonic field 
\begin{align*}
\3X\rt=\3A\orr\cos(\o_0 t+\vf),\qq\o_0>0,
\end{align*}
\eq{Xo} and \eq{X0} give
\begin{align}\lab{time-harm}
\mb X_\o\orr=2\p\3A\orr e^{i\vf}\d(\o-\o_0)\ \ \ \ \text{and}\ \ \ \ \mb X\rt=2\3A\orr e^{i\vf} e^{i\o_0 t}.
\end{align}
This shows that \eq{X00} is a generalization of the \it phasor representation \rm to fields with general time dependence. 

Since $\o$ is positive in the integral \eq{X0}, there is no harm in complexifying $t$ as long as its imaginary part is \it positive. \rm The formal substitution
\begin{align}\lab{tau}
t\to\t\=t+is,\ s>0
\end{align}
gives
\begin{align}\lab{X}
\mb X'(\3r,\t)=\frac1\p \int_0^\infty\dd\o\,e^{i\o\t}\3X_\o'\orr
=\frac1\p \int_0^\infty\dd\o\,e^{i\o t}e^{-\o s}\3X_\o'\orr,
\end{align}
and \eq{Xo} extends to $s>0$ as
\begin{align}\lab{Xo1}
\mb X_\o'(\3r, s)\=\ir\dd t\,e^{-i\o t}\mb X'(\3r, t+is)=2H\0\o e^{-\o s}\3X_\o'\orr.
\end{align}
The factor $e^{-\o s}$ acts as a \it filter \rm suppressing high-frequency components, thus \it smoothing \rm the field in the time domain. Moreover, this occurs smoothly in the \it frequency \rm domain, thus avoiding anomalies like the Gibbs phenomenon in the time domain. The larger $s$, the smoother $\mb X'$ becomes.

This shows that the role of $s$ in the frequency domain is to suppress high frequencies.  To see its role in the time domain, note that
\begin{align}\lab{C0}
\3X\rt=\d\0t\3X^0\orr\imp\3X_\o\orr\=\3X^0\orr  \imp \mb X(\3r,\t)=C\0\t\,\3X^0\orr,
\end{align}
where
\begin{align}\lab{C}
C\0\t=\frac1\p \int_0^\infty\dd\o\, e^{i\o\t}=\frac i{\p\t}=\frac{s+it}{\p(s^2+t^2)}
\end{align}
is called the \it Cauchy kernel. \rm Equation \eq{C0} shows the role of $s$ in the time domain:
Since even an \it impulse \rm $\d\0t$ in $\3X'$ results in a pulse $C(t+is)$ of width $\D t\sim s$ in $\mb X'$, 
\begin{align}\lab{dt>s}
{\blue \mb X'(\3r,t+is)\ \text{\it cannot contain spikes in $t$ narrower than $\5O\0s$,}}
\end{align}
which can be expressed roughly as $\D t>s$. Thus $s$ is a \it time-resolution scale, \rm or \it scale \rm for short. We will call $\mb X'(\3r, t+is)$ the \it time-scale representation \rm of $\3X'$. It will turn out to represent $\3X$ at all scales $s'\ge s$ (Remark \ref{R:cumulative}).

Equation \eq{X} suggests that $\mb X'$ is \it analytic \rm in $\t\in\4C\9+$, the upper-half complex time plane. This depends, of course on the behavior of $\3X_\o'$ as a function of $\o$. The complex time derivative of \eq{X} is
\begin{align}\lab{Xt0}
\pl_\t\mb X'(\3r,\t)=\frac i\p \int_0^\infty\dd\o\,\o e^{i\o\t}\3X_\o'\orr,
\end{align}
\it provided \rm the integral converges. A sufficient (but not \it necessary\rm) condition for analyticity is that $\3X'$ be \it square-integrable,\,\rm\footnote{Obviously $\3X\notin L^2$ if $\3X\dc\ne\30$, which is why we have defined $\3X'=\3X-\3X\dc$ in \eq{XX'}. 
}
written $\3X'\in L^2$:
\begin{align}\lab{L2t}
\|\3X'\|^2\=\ir\dd t\,|\3X'\rt|^2<\infty\ \ \hbox{for all}\ \ \3r.
\end{align}
This condition will be assumed throughout.\footnote{In the engineering literature, square-integrable fields are sometimes said to have \it finite energy. \rm Since we will be computing physical energies, this terminology will be avoided.
}
By \eq{Xo1} and the \it Plancherel identity \rm (which states that $\3X'$ has the same $L^2$ norms in the time domain and the frequency domain), \eq{L2t} becomes
\begin{align}\lab{L2}
\|\3X'\|^2=\frac1{2\p}\ir\dd\o\,|\3X_\o'\orr|^2=\frac1{\p}\int_0^\infty\dd\o\,|\3X_\o'\orr|^2<\infty,
\end{align}
where \eq{rc} was used in the last equality.
Define the \it $s$-norm \rm of $\mb X'$ for fixed $s\ge 0$ (and a corresponding inner product $\la\mb X',\mb Y'\ra_s$) by
\begin{align}\lab{snorm}
\|\mb X'\|_s^2\=\ir\dd t\,|\mb X'(\3r,t+is)|^2=\frac2\p\int_0^\infty\dd\o\, e^{-2\o s} |\3X_\o'\orr|^2,
\end{align}
where we used \eq{Xo1}. Note that for $s=0$, this is related to the norm \eq{L2} by\footnote{The factor 2 on the right will be explained in Remark \ref{R:RMS}.
}
\begin{align}\lab{snorm0}
\|\mb X'\|_0^2=2\|\3X'\|^2.
\end{align}
If $\3X'\in L^2$, then \eq{Xt0} and \eq{snorm} give
\begin{align}\lab{Xt2}
\|\pl_\t\mb X\|_s^2=\ir\dd t\,|\pl_\t\mb X|^2=\frac2\p\int_0^\infty\dd\o\, \o^2\,e^{-2\o s}|\3X_\o'\orr|^2<\infty
\end{align}
for all $\3r$ and all $s>0$, since $\o^2\,e^{-2\o s}$ is bounded. This proves that $\mb X'$ is defined as an analytic function of $\t$ in $\4C\9+$ (hence the name `analytic signal').

\rem\rm\label{R:s>0}\label{R:analyticity}
In physics and engineering, analytic signals are usually restricted to their $s=0$ \it boundary values, \rm as in \eq{X0}. However, their analyticity and the physical interpretation of $s>0$ will be seen to play a critical role here.

Let us compute the real and imaginary parts of $\mb X'$. By \eq{X} and \eq{rc}, the complex conjugate of $\mb X'(\3r,\t)$ is the function of $\7\t=t-is$ given by
\begin{align*}
\bmb X'(\3r,\7\t)=\frac1\p \int_0^\infty\dd\o\,e^{-i\o\7\t}\bb X_\o'\orr
=\frac1\p \int_{-\infty}^0\dd\o\,e^{i\o\7\t}\3X_\o'\orr,
\end{align*}
which is an analytic continuation of the \it negative-frequency part \rm of $\3X'$ to $\7\t\in\4C\9-$. Therefore the real and imaginary parts of $\mb X'(\3r,\t)$ are\footnote{By \eq{X00}, $\mb X_1=\3X\dc+\mb X_1'$ and $\mb X_2=\mb X_2'$ since $\3X\dc$ is real, so there is no need to use $\mb X_2'$.
}
\begin{equation}\lab{X121}\begin{split}
\mb X_1'(\3r, t, s)&=\frac12(\mb X'+\bmb X')=\frac1{2\p }\ir\dd\o\,e^{-|\o|s}e^{i\o t}\3X_\o'\orr\\
\mb X_2(\3r, t, s)&=\frac1{2i}(\mb X'-\bmb X')=\frac1{2i\p}\ir\dd\o\,\sgn\o\,e^{-|\o|s}e^{i\o t}\3X_\o'\orr,
\end{split}\end{equation}
where $\sgn\o=\pm 1$ is the sign of $\o$. 
As $s\to0$, the $s$-norm of $\mb X_1'$ goes over continuously to the time-domain norm:
\begin{align}\lab{snorm1}
s\to0\imp\mb X_1'\to\3X'\ \ \text{and}\ \  \|\mb X_1'\|_s\to\|\3X'\|.
\end{align}
Equations \eq{X121} show that $\mb X_1'$ and $\mb X_2$ are orthogonal with equal norms:
\begin{align}\lab{ON}
\|\mb X'_1\|_s^2=\|\mb X_2\|_s^2\, \ \ \text{and}\ \ \la\mb X'_1,\mb X_2\ra_s=0.
\end{align}

\rem\rm\label{R:RMS} By \eq{snorm1} and \eq{ON}, we have
\begin{align}\lab{rms0}
\|\mb X'\|_0^2=\|\mb X'_1+i\mb X_2\|_0^2=\|\mb X_1'\|_0^2+\|\mb X_2\|_0^2=2\|\3X'\|^2.
\end{align}
This explains the factors of 2 in \eq{time-harm} and \eq{snorm0}. $\mb X'$ is normalized so that $\Re\mb X'\to\3X'$ as $s\to 0$.  Hence $\|\3X'\|$ is the \it root mean square \rm (RMS) of $\|\mb X_1'\|_{s=0}$ and $\|\mb X_2\|_{s=0}$,
\begin{align}\lab{rms}
\|\3X'\|=\sr{\frac{\|\mb X_1'\|_0^2+\|\mb X_2\|_0^2}2}=\frac{\|\mb X'\|_0}{\sr{2}}.
\end{align}
Consequently, the coefficients of \it scaled quadratic variables \rm such as \eq{SUP} will be \it half \rm the coefficients of their \it local \rm counterparts \eq{USP}. For example, the local Poynting vector is $\3E\times\3H$ and its scaled counterpart is $\tfrac12\mb E\times\bmb H$.

\rem\rm\label{R:equivalent}
Equation \eq{snorm1} shows that $\3X$ can be recovered from $\Re\mb X$ \it continuously \rm in the norms. The time-scale representation $\mb X(\3r, \t)$ is therefore completely equivalent to the time representation $\3X(\3r, t)$ and the frequency representation $\3 X_\o\orr$.  Each has its own qualities, and it cannot be said that one is more `real' than the others. It turns out that $\mb X$ is useful for analyzing reactive energy.

\section{The analytic-signal transform}\label{S:ast}

We can express the transformation $\3X\rt\to\mb X(\3r,\t)$ directly in the time domain by using the identity
\begin{align}\lab{H}
\frac1{i\p}\ir\frac{\dd t'}{t'-\t}\,e^{i\o t'}=2H\0\o e^{i\o\t}\ \ \hbox{for all}\ \  \o\ne0\ \ \hbox{and}\ \ \t\in\4C\9+.
\end{align}
This is obtained from Cauchy's residue theorem by closing the contour in the upper half-plane if $\o>0$, and in the lower half-plane if $\o<0$. Substituting \eq{H} into \eq{X} and reversing the order of integration (justified if $\3X'\in L^2$ since the integrals converge absolutely) gives
\begin{align}\lab{ast}
\mb X'(\3r,\t)=\frac i\p \ir\frac{\dd t'}{\t-t'}\,\3X'(\3r, t').
\end{align}
Thus $\mb X'$ is a convolution
\begin{align}\lab{ast1}
\mb X'(\3r,t+is)=\ir\dd t'\,C_s(t-t')\3X'(\3r, t')\=C_s*\3X'\rt,
\end{align}
where 
\begin{align}\lab{CPH}
C_s\0t=\frac i{\p (t+is)}=\fr P_s\0t+i\fr H_s\0t
\end{align}
is the Cauchy kernel already encountered in \eq{C}, with 
\begin{align}\lab{PH}
\fr P_s\0t=\frac s{\p(s^2+t^2)}\ \ \text{and}\ \  \fr H_s\0t=\frac t{\p(s^2+t^2)},\ \ s>0.
\end{align}
The integral \eq{ast} is called a \it analytic-signal transform \rm \ci{K3, K11}. Note that\footnote{The second integral is defined by integrating from $-b$ to $b$ and then letting $b\to\infty$.
}
\begin{align}\lab{PH1}
\fr P_s\0t>0,&& \ir\dd t\  \fr P_s\0t=1,&& \ir\dd t\, \fr H_s\0t=0.
\end{align}
$\fr P_s$ is known as the \it Poisson kernel. \rm 
The time-domain versions of \eq{X121} are
\begin{equation}\lab{X120}\begin{split}
\mb X_1'(\3r,t,s)&=\frac1\p \ir\frac{s\,\dd t'}{s^2+(t-t')^2}\,\3X'(\3r, t')\\
\mb X_2(\3r,t,s)&=\frac1\p \ir\frac{(t-t')\dd t'}{s^2+(t-t')^2}\,\3X'(\3r, t'),
\end{split}\end{equation}
or briefly
\begin{align}\lab{X12}
\mb X_1'=\fr P_s*\3X'\ \ \text{and}\ \  \mb X_2=\fr H_s*\3X'.
\end{align}
In the limit $s\to 0$ we have
\begin{align*}
\fr P_s\0t\to\d\0t,\ \ \hbox{hence}\ \  \mb X_1'(\3r,t,s)\to\3X'\rt,
\end{align*}
as already noted in \eq{snorm1}. Furthermore, as $s\to0$,
\begin{align*}
\mb X_2(\3r, t,s)\to\frac1{\p}\ppint_{-\infty}^\infty\frac{\dd t'}{t-t'}\,\3X'(\3r, t')\=\fr H\3X'\rt,
\end{align*}
where $\ppint$ is the \it principal-value integral \rm and $\fr H$ is the \it Hilbert transform. \rm
Thus
\begin{align}\lab{Xs0}
s\to 0\imp\mb X'(\3r,t+is)\to\3X'\rt+i\fr H\3X'\rt.
\end{align}
Note that \eq{Xs0} can be read off directly from \eq{Xo} since $2H\0\o=1+\sgn\o$ and \eq{X121} shows that
\begin{align}\lab{HT}
(\fr H\3X')_\o\orr=-i\sgn\o\,\3X'_\o\orr.
\end{align}

\rem\rm\label{R:laglead}
Equation \eq{HT} can be rewritten as
\begin{align}
e^{i\o t}(\fr H\3X')_\o\orr&=\begin{cases}
e^{i(\o(t-\p/2\o))}\3X'_\o\orr, &\o>0\\ e^{i(\o(t+\p/2\o))}\3X'_\o\orr, &\o<0
\end{cases}\nt\\
&=e^{i(\o(t-\p/2|\o|))}\3X'_\o\orr,\qq  \o\ne 0. \lab{lag}
\end{align}
Thus $\fr H$ performs a frequency-independent \it phase lag \rm by $\p/2$ or, equivalently, a frequency-dependent \it time lag \rm by a quarter period $\p/2|\o|$ on each Fourier component. 

\rem\rm\label{R:fluctuations} 
Equation \eq{X12} shows that 
\begin{equation}\lab{X123}\begin{split}
&\mb X_1'(\3r, t,s)\  \hbox{is $\3X'\rt$ \ \it smoothed in time to the scale \rm $s$\rm}\\
&\mb X_2(\3r, t,s)\  \hbox{is $\fr H\3X'\rt$ \it smoothed in time to the scale \rm $s$,\rm}
\end{split}\end{equation}
or briefly that $(\mb X_1', \mb X_2)$ are \it scaled versions \rm of $(\3X', \fr H\3X')$. From \eq{PH1} it follows that
\begin{align}\lab{fluct0}
\ir\dd t\,\mb X_2(\3r, t+is)=\30\ \ \text{and}\ \  C_s*\3X\dc\orr=\3X\dc\orr.
\end{align}
Therefore $\mb X_2$ can be interpreted as a \it zero-mean fluctuation, \rm and Equation \eq{ast1} extends to $\3X=\3X\dc+\3X'$:
\begin{align}\lab{ast2}
\mb X(\3r,t+is)=C_s*\3X\rt.
\end{align}

\rem\rm\label{R:timescale}
In the analytic-signal transform \eq{ast1}, note that 
\begin{align}\lab{ctsa}
C_s(t-t')=s\inv C_1\lp\frac{t-t'}s\rp,
\end{align}
which is typical of \it continuous time-scale analysis.\rm\footnote{The factor $s^{-1/2}$ is often used instead of $s\inv$, but this is just convention \ci[p.~67]{K11}.
}

\rem\rm\label{R:sto8} Writing $\mb X'$ in \eq{X} as a vector-valued inner product in the positive-frequency domain,
\begin{align}\lab{Xip}
\mb X'(\3r, t+is)=\frac1\p \int_0^\infty\dd\o\,e^{-\o s} e^{i\o t}\3X_\o'\orr\=\la e^{-\o s}, e^{i\o t}\3X_\o'\orr\ra,
\end{align}
it follows from the Cauchy-Schwarz inequality that
\begin{align}\lab{Xs}
|\mb X'(\3r, t+is)|^2\le\|e^{-\o s}\|^2 \|\3X_\o'\|^2=\frac1{2\p s}\|\3X'\|^2,
\end{align}
where we have used \eq{L2}. Thus $\3X\dc$ is a \it boundary value \rm of $\mb X$ in the sense that
\begin{align}\lab{Xto0}
s\to\infty\imp\mb X(\3r,t+is)\to \3X\dc\orr,
\end{align}
which explains the notation $\3X\dc$. Furthermore, by \eq{X},
\begin{align*}
-\pl_s|\mb X(\3r, \t)|^2
&=\frac1{\p^2}\int_0^\infty\dd\o\int_0^\infty\dd\o'\,(\o+\o')[e^{i\o\t}\3X_\o'],
\cdot\overline{[e^{i\o'\t}\3X_{\o'}]}
\end{align*}
and the right side can be shown to be nonnegative. Hence
\begin{align}\lab{Xs1}{\blue 
-\pl_s|\mb X(\3r, t+is)|^2\ge 0\ \ \text{for all $\rt$ and $s>0$.}}
\end{align}
Equations \eq{Xto0} and \eq{Xs1} together prove the following.

\rem\label{R:monotonic}  \bf (static limit): \rm  As $s\to\infty, \mb X(\3r,t+is)\to \3X\dc\orr$ \it monotonically. \rm

This will be used to investigate the scale dynamics of reactive energy \eq{tune}.

\rem\rm\label{R:nonloc}
Equations \eq{ast} and \eq{X120} show that the analytic-signal  transform is \it non-local: \rm if a current density $\3J\rt$ vanishes outside some time interval $[t_1, t_2]$, its analytic signal $\mb J(\3r, t+is)$ need \it not \rm vanish for $t\notin [t_1, t_2]$. Instead, its real part $\mb J_1$ vanishes in the limit $s\to0$ when $t\notin[t_1, t_2]$, since $\mb J_1\to\3J$. The imaginary part $\mb J_2$, however, does not vanish in that limit since $\fr H_s\0t\nrightarrow 0$. In fact, \eq{PH} tells us that $\mb J_2$ is  \it even less local \rm than $\mb J_1$ because
\begin{align}\lab{tinv}
\fr P_s\0t=\frac s{\p(s^2+t^2)}=\5O(t^{-2})\ \ \hbox{and}\ \   \fr H_s=\frac t{\p(s^2+t^2)}=\5O(t\inv).
\end{align}
But \eq{fluct0} shows that $\mb J_2$ is a \it zero-mean fluctuation, \rm \ie
\begin{align}\lab{fluct}
\ir\dd t\,\mb J_2(\3r, t+is)=\30.
\end{align}
A rough measure of the \it degree \rm of non-locality is given by $\D t\sim s$. For this reason we refer to $\3X$ as a \it local \rm variable and $\mb X$ as a  \it scaled \rm variable.

\section{Conservation of active and reactive energy}\label{S:cons}

We now apply the above machinery to a general electromagnetic field $(\3E,\3H)$ in vacuum with charge-current density $(\r,\3J)$, so that $\mb E,\mb H,\mb J,\uprho$ are analytic in $\t\in\4C\9+$.  To keep the notation simple and focus on the essentials, we use \it natural Heaviside-Lorentz units,\rm\footnote{However, we will reinsert factors of $c$ when they help clarify a statement.
}
where $\e_0=\m_0=c=1$ and Maxwell's equations are
\begin{align}\lab{M0}
\curl\3E&=-\pl_t\3H&& \curl\3H=\pl_t\3E+\3J\\
\div\3E&=\r&&\ \ \div\3H=0.\nt
\end{align}
Since $t=(\t+\7\t)/2$ and $s=i(\7\t-\t)/2$, it follows that
\begin{align}\lab{plt}
\pl_\t&\=\frac{\pl}{\pl\t}=\tfrac12(\pl_t-i\pl_s) && \7\pl_\t\=\frac{\pl}{\pl\7\t}=\tfrac12(\pl_t+i\pl_s)
\end{align}
and
\begin{align}\lab{plt2}
\pl_t=\7\pl_\t+\pl_\t && i\pl_s=\7\pl_\t-\pl_\t.
\end{align}
If $\mb X$ is analytic in $\t$, then $\7\pl_\t\mb X=\pl_\t\bmb X=\30$, hence
\begin{align}\lab{pls}
\pl_\t\mb X=\pl_t\mb X=-i\pl_s\mb X\ \ \text{and}\ \  \7\pl_\t\bmb X=\pl_t\bmb X=i\pl_s\bmb X.
\end{align}
Applying this to \eq{M0} gives the \it analytic Maxwell equations \rm
\begin{align}\lab{M}
\curl\mb E&=-\pl_\t\mb H &&\curl\mb H=\pl_\t\mb E+\mb J\\
\div\mb E&=\uprho &&\ \ \div\mb H=\30.\nt
\end{align}
To generalize Poynting's theorem, start as usual with the identity
\begin{align*}
\div(\mb E\times\bmb H)&=\bmb H\cdot\curl\mb E-\mb E\cdot\curl\bmb H.
\end{align*}
By \eq{M}, this gives
\begin{align}\lab{cpt0}
\pl_\t\mb H\cdot\bmb H+\mb E\cdot\7\pl_\t\bmb E+\div(\mb E\times\bmb H)=-\mb E\cdot\bmb J.
\end{align}
Since $\pl_\t\bmb H=0$ and $\7\pl_\t\mb E=0$, this is equivalent to
\begin{align*}
\pl_\t|\mb H|^2+\7\pl_\t|\mb E|^2+\div(\mb E\times\bmb H)=-\mb E\cdot\bmb J.
\end{align*}
Inserting the expressions \eq{plt} for $\pl_\t$ and $\7\pl_\t$ and collecting terms gives
\begin{align}\lab{P0}
\tfrac12\pl_t\lp|\mb H|^2+|\mb E|^2\rp-\frac i2\pl_s\lp |\mb H|^2-|\mb E|^2\rp+\div(\mb E\times\bmb H)
=-\mb E\cdot\bmb J.
\end{align}
Now define the variables
\begin{align}
&\5W_m(\3r, t,s)=\tfrac14|\mb H|^2&&\5W_e(\3r, t,s)=\tfrac14|\mb E|^2\nt\\
&\5U(\3r,t,s)=\5W_m+\5W_e && \5X(\3r,t,s)=\5W_m-\5W_e \nt\\
&\mb S(\3r,t,s)=\tfrac12\Re\,(\mb E\times\bmb H)&& \mb T(\3r,t,s)=\tfrac12\Im\,(\mb E\times\bmb H)\lab{SUP}\\
&\5P(\3r,t,s)=-\tfrac12\Re\,(\mb E\cdot\bmb J) && \5Q(\3r,t,s)=-\tfrac12\Im\,(\mb E\cdot\bmb J).\nt
\end{align}
Then \eq{P0} becomes the \it complex Poynting theorem in the time-scale domain \rm
\begin{align}\lab{CPT}{\blue 
\pl_t\,\5U-i\pl_s\5X+\tfrac12\div(\mb E\times\bmb H)=-\tfrac12\mb E\cdot\bmb J,}
\end{align}
which splits into the \it active and reactive energy conservation laws \rm  
\begin{gather}{\blue 
\ \pl_t\,\5U+\div\mb S=\5P\ } \lab{P}\\
{\blue \!\!-\pl_s\5X+\div\mb T\!=\5Q.\!\!}\lab{Q}
\end{gather}
The variables \eq{SUP} have the following physical interpretations:
\begin{align*}
&\5W_m(\3r, t, s)\ \hbox{is a scaled version of the magnetic energy density $W_m\rt$}\\
&\5W_e(\3r, t, s)\ \hbox{is a scaled version of the electric energy density $W_e\rt$}\\
&\5U(\3r, t, s)\ \hbox{is a scaled version of the active energy density $U\rt$}\\
&\5X(\3r, t, s)\ \hbox{is a scaled version of the reactive energy density $X\rt$}\\
&\mb S(\3r, t, s)\ \hbox{is a scaled version of the active energy flux density $\3S\rt$}\\
&\mb T(\3r, t, s)\ \hbox{is a scaled reactive energy flux density}\\
&\5P(\3r, t, s)\ \hbox{is a scaled version of the active power density $P\rt$}\\
&\5Q(\3r, t, s)\ \hbox{is a scaled reactive power density,}
\end{align*}
where
\begin{align}\lab{Wme}
W_m=\tfrac12\3H^2\ \ \text{and}\ \ W_e=\tfrac12\3E^2
\end{align}
are the \it local \rm magnetic and electric energy densities of the field,
\begin{align}\lab{USP}
U=W_m+W_e, && \3S=\3E\times\3H, && P=-\3E\cdot\3J
\end{align}
are the local energy density, Poynting vector and power density satisfying the \it local Poynting theorem \rm \ci{J99}
\begin{align}\lab{RP}
\pl_t U+\div\3S=P.
\end{align}

Note that we have the scaled \it complex \rm power flux and power density
\begin{align}\lab{cx1}
\mb S+i\mb T=\tfrac12\mb E\times\bmb H &&\5P+i\5Q=-\tfrac12\mb E\cdot\bmb J,
\end{align}
but \it there is no scaled complex energy density; \rm see Remark \ref{R:noCxenergy}. 

\rem\rm\label{R:new}
Carozzi, Bergman and Karlsson \ci{CBK5} also studied the generalization of the complex Poynting theorem to analytic signals, but they restricted their analysis to $s=0$ and thus had no recourse to analyticity. Instead of \eq{cpt0}, they obtained
\begin{align*}
\bmb H\cdot\pl_t\mb H+\mb E\cdot\pl_t\bmb E+\div(\mb E\times\bmb H)=-\mb E\cdot\bmb J,
\end{align*}
where $\mb E=\mb E(\3r,t)$, etc., are the analytic signals with $s=0$. They then proved that this \it cannot \rm lead to a conservation law for reactive energy in $t$. Indeed, as we have seen, reactive energy is not conserved in time but in \it scale, \rm which will be interpreted as \it reactive time \rm in Section \ref{S:interpretation}.

\rem\rm\label{norad} The \it far fields \rm ($r\to\infty$) satisfy 
\begin{align}\lab{far}
\mb H_{\rm far}=\bh r\times\mb E_{\rm far} &&\mb E_{\rm far}=-\bh r\times\mb H_{\rm far},&& 
\end{align}
hence
\begin{align}\lab{far1}
&|\mb H_{\rm far}|^2=|\mb E_{\rm far}|^2 &&  \mb E_{\rm far}\cdot\mb H_{\rm far}=0&& \5X_{\rm far}=0\\
&\mb E_{\rm far}\times\bmb H_{\rm far}=\bh r\,|\mb E_{\rm far}|^2 &&
\mb S_{\rm far}=\tfrac12\bh r\, |\mb E_{\rm far}|^2 && \mb T_{\rm far}=\30.\nt
\end{align}
As expected, \it reactive energy does not radiate and vanishes in the far zone. \rm 

\rem\rm\label{R:inertia}
In \ci{K11a, K12} I defined the \it electromagnetic inertia density\,\rm\footnote{Since $c\equiv 1$, $\3S\rt$ is the field's \it momentum density \rm and  $I\rt$ is the field analogue of the \it rest energy \rm (or mass) of a relativistic particle.
}
\begin{align}\lab{I}
I\rt\=\sr{U^2-\3S^2}=\sr{\tfrac14(\3H^2-\3E^2)^2+(\3E\cdot\3H)^2}
\end{align}
and showed that it represents field energy that is \it locally at rest. \rm  Note that $I\rt$ vanishes in the far zone
due to the local version of \eq{far1}. For this reason, I believed at the time that \eq{I} could represent the field's reactive energy density. But it is now clear that the condition of being locally \it at rest \rm is too strict. To see how the two concepts are related,  compute the scaled version of $I\rt$:
\begin{align}\lab{5I}{\blue 
\5I(\3r, t, s)\=\sr{\5U^2-\mb S^2}=\sr{\5X^2+\5R^2},}
\end{align}
where $\5X$ is defined in \eq{SUP} and
\begin{align}\lab{mom}{\blue 
\5R(\3r, t,s)=\tfrac12 |\mb E\cdot\mb H|.}
\end{align}
The local versions of $\5X$ and $\5R$,
\begin{align}\lab{RX}
X\rt=\frac12(\3H^2-\3E^2)\ \ \ \ \text{and}\ \ \ \ R\rt=|\3E\cdot\3H|,
\end{align}
are the two fundamental \it Lorentz scalars \rm of the field.\footnote{$\3E\cdot\3H$ is a \it pseudoscalar \rm since it changes sign upon spatial reflection,  which is why we must take its absolute value to obtain a scalar.
}
Then \eq{5I} shows that
\begin{align}\lab{EH0}
\5I\ge |\5X|, \ \ \ \text{and}\ \ \ \5I=|\5X|\iff \5R=0.
\end{align}
$\5X$ and $\5R$ are two forms of internal energy contributing symmetrically to $\5I$.

\rem\rm\label{R:rms}
As explained in Remark \ref{R:RMS}, the local variables \eq{USP} have \it twice \rm the coefficients of their scaled versions \eq{SUP}. 

\rem\rm\label{R:CPT} It is instructive to see how \eq{CPT} reduces to the usual complex Poynting theorem for time-harmonic fields. If the analytic signals have just one Fourier component with $\o>0$ (which rules out DC components), then
\begin{align*}
\mb E(\3r,\t)=e^{i\o\t}\mb E_o\orr, && \mb H(\3r,\t)=e^{i\o\t}\mb H_o\orr, && \mb J(\3r,\t)=e^{i\o\t}\mb J_o\orr
\end{align*}
and
\begin{align*}
|\mb E|^2&=e^{-2\o s}|\mb E_o\orr|^2, && |\mb H|^2=e^{-2\o s}|\mb H_o\orr|^2\\
\mb E\times\bmb H&=e^{-2\o s}\mb E_o\times\bmb H_o,&&\mb E\cdot\bmb J=e^{-2\o s}\mb E_o\cdot\bmb J_o.
\end{align*}
In this special case, all the densities in \eq{SUP} are time-independent and can thus be identified as averages over the period $T=2\p/\o$.  Equation \eq{CPT} becomes
\begin{align*}
\pl_t\,\5U-i\pl_s\5X+\tfrac12e^{-2\o s}\div(\mb E_o\times\bmb H_o)=-\tfrac12e^{-2\o s}\mb E_o\cdot\bmb J_o.
\end{align*}
Since 
\begin{align}\lab{Rs}
\pl_t\,\5U=0\qq\ \ \hbox{and}\ \ \qq -2i\pl_s\5X=4i\o\5X=i\o\, e^{-2\o s}(|\mb H_o|^2-|\mb E_o|^2),
\end{align}
this reduces to the usual time-harmonic complex Poynting theorem
\begin{align}\lab{CP}
i\o (|\mb H_o|^2-|\mb E_o|^2)+\div(\mb E_o\times\bmb H_o)=-\mb E_o\cdot\bmb J_o.
\end{align}
The first term thus comes from $-i\pl_s\5X$.  Although \eq{CP} can be used to \it define \rm $\5X$ in the time-harmonic case, this cannot be \it interpreted \rm in terms of reactive energy without the conservation law \eq{Q}, and it certainly cannot be extended to the time domain.  That is why discussions of the imaginary part of \eq{CP} usually dance around the issue of reactive energy, as mentioned at the beginning of Section \ref{S:intro}.\footnote{An indirect and merely \it qualitative \rm connection with reactive energy is made by showing that the first term in \eq{CP} is related to the \it reactance \rm  of a circuit or an antenna \ci[Section 12.5]{HMZS8}.
}

\rem\rm\label{R:TQ2} While an expression for reactive energy density has been missing, the complex Poynting theorem \eq{Q} has been used to define the reactive energy flux density $\mb T$ and the power density $\5Q$ for time-harmonic fields. However, even these definitions are \it incomplete \rm without a proper expression for reactive energy. Furthermore, the conservation law \eq{Q} implies that $\mb T$ and $\5Q$ are rates not with respect to $t$, as is usually assumed, but with respect to $s$ (Section \ref{S:rtime}). This has resulted in an \it ad hoc \rm practice of measuring reactive power in \it volt-ampere reactive \rm [VAr] units rather than watts, which cannot be fully justified until the above connection is understood.

\rem\rm\label{R:static} By Remark \ref{R:monotonic} , the \it static limits \rm $s\to\infty$ of \eq{SUP} are
\begin{align}
\5U&\to\tfrac14(\3H\dc^2+\3E\dc^2)=\tfrac12U\dc\orr, 
&& \5X\to\tfrac14(\3H\dc^2-\3E\dc^2)=\tfrac12X\dc\orr\nt\\
\mb S&\to\tfrac12\3E\dc\times\3H\dc=\tfrac12\3S\dc\orr&& \mb T\to\30\lab{lims}\\
\5P&\to-\tfrac12\3E\dc\cdot\3J\dc=\tfrac12P\dc\orr,&& \5Q\to\30.\nt
\end{align}
\it The variables \eq{SUP} converge monotonically to half their local static values. \rm

The factors  $\tfrac12$ in \eq{lims} are harmless. They mean simply that in the static limit we leave the world of analytic signals and return to the `RMS world' of local fields (Remark \ref{R:RMS}). The active energy conservation law \eq{P} reduces to
\begin{align}\lab{Ps0}
\div\3S\dc=P\dc,
\end{align}
and the reactive one \eq{Q} reduces to the identity $0=0$.

\rem\rm\label{R:modes}
To illustrate how scaling works, suppose the fields consist of $n$ isolated \it modes \rm with (not necessarily harmonic) frequencies
\begin{align*}
0<\o_1<\o_2< \cdots<\o_n.
\end{align*}
Then the densities in \eq{SUP} are governed by the mode factors
\begin{align*}
e^{i\o_k\t-i\o_l\7\t}=e^{i(\o_k-\o_l)t}e^{-(\o_k+\o_l)s},\qq k\ne l,
\end{align*}
representing \it scale-filtered beats, \rm as well as static terms ($k=l$) representing period averages. If we begin at a coarse scale $s\gg 1/(2\o_1)$, where all the densities are negligible, and slowly \it refine \rm $s$, then higher and higher modes emerge one by one in the order of increasing $\o_k+\o_l$. When $0<s < 1/(2\o_n)$, all the modes have emerged. Further scale refinements cannot resolve new modes, only improve the resolution of the modes resolved earlier. 

\rem\rm\label{WmWe}
According to \eq{Xs1},
\begin{align}\lab{tune}{\blue 
-\pl_s\5W_m\ge 0\ \ \text{and}\ \ -\pl_s\5W_e\ge 0.}
\end{align}
Hence the magnetic and electric energies both grow as the scale gets finer.

\section{Can reactive power be instantaneous?}\label{S:s=0}

This question has been debated in the power engineering community for more than one hundred years, and to this day there has been no satisfactory resolution \ci{E10}. There are a number of conflicting models, each with its own way of measuring, metering and regulating reactive energy and power, but none is based on universal physical principles. The reason is simple: there is no fundamental theory of reactive energy, hence the different models form a patchwork of \it ad hoc \rm constructions, each with its pros and cons. The absence of an underlying physical theory has economic consequences, putting strains on the power grid resulting in inefficiency, overheating, and the necessity to overbuild in order to protect the system.

We now show that  \it  according the conservation theorem \eq{Q}, no local value may be assigned to the reactive energy density $\5X$, its flux $\mb T$, and its power density $\5Q$ \rm --- even in the sharp-time limit $s=0$.

Assume, for simplicity, that our fields $\3H,\3E,\3J$ have no DC components. Then \eq{X121} gives the decompositions
\begin{align}\lab{sto0}
\mb X(\3r, t,s)=(\fr P_s\*\3X)\rt+i(\fr H_s\*\3X)\rt\=\mb X_1(\3r, t,s)+i\mb X_2(\3r, t,s)
\end{align}
for  $\mb E, \mb H, \mb J$. Recall that the scaled Poisson and Hilbert kernels $\fr P_s$ and $ \fr H_s$ decay as $t^{-2}$ and $t^{-1}$, respectively, and that 
\begin{align*}
\lim_{s\to0} \mb X_1(\3r, t,s)=\3X\rt.
\end{align*}
A function of $(\3r, t, s)$ depending only on $(\mb E_1, \mb H_1, \mb J_1)$ will be called \it semi-local \rm because it becomes local (instantaneous) in the limit $s\to 0$, and a function containing any of the fields $(\mb E_2, \mb H_2, \mb J_2)$ will be called \it nonlocal. \rm  Substituting the decompositions \eq{sto0} into \eq{SUP} and suppressing the variables $(\3r, t,s)$ gives 
\begin{align*}
&\5U=\5U_1+\5U_2, && \5U_1\=\tfrac14\mb H_1^2+\tfrac14\mb E_1^2, && \5U_2\=\tfrac14\mb H_2^2+\tfrac14\mb E_2^2\\
&\mb S=\mb S_1+\mb S_2, &&\mb S_1\=\tfrac12\mb E_1\times\mb H_1,&& \mb S_2\=\tfrac12\mb E_2\times\mb H_2\\
&\5P=\5P_1+\5P_2, && \5P_1\=-\tfrac12\mb E_1\cdot\mb J_1, && \5P_2\=-\tfrac12\mb E_2\cdot\mb J_2
\end{align*}
and
\begin{equation}\lab{XTQ}\begin{split}
&\5X=\tfrac14\!\lp\mb H_1^2-\mb E_1^2\rp+\tfrac14\!\lp\mb H_2^2-\mb E_2^2\rp\=\5X_1+\5X_2\\
&\mb T=\tfrac12\!\lp\mb E_2\times\mb H_1+\mb E_1\times\mb H_2\rp\\
&\5Q=\tfrac12\!\lp\mb E_1\cdot\mb J_2-\mb E_2\cdot\mb J_1\rp.
\end{split}\end{equation}
The \it active \rm quantities $\5U, \mb S,\5P$ thus split neatly into semi-local parts $\5U_1, \mb S_1,\5P_1$ and nonlocal parts  $\5U_2, \mb S_2,\5P_2$. Since the semi-local and the nonlocal parts of $\mb E, \mb H, \mb J$ each satisfy Maxwell's equations for any fixed $s\ge 0$, the \it real \rm Poynting theorem \eq{RP} requires that their active energies be conserved independently:
\begin{align}\lab{P12}
\pl_t \,\5U_k+\div\mb S_k=\5P_k, \ \ k=1,2.
\end{align}
The active-energy Poynting theorem \eq{P} thus splits into separate conservation laws for the semi-local and the nonlocal parts of the energy. In the limit $s=0$, the semi-local part of \eq{P12} reduces to the real Poynting theorem \eq{RP} for $U, \3S, P$, with
\begin{align*}
\5U_1(\3r,t,0)=\tfrac12 U\rt, && \mb S_1(\3r,t,0)=\tfrac12\3S\rt, && \5P_1(\3r, t, 0)=\tfrac12 P\rt.
\end{align*}
Thus we have obtained a smooth connection between the real part \eq{P} of \eq{CPT} and the real Poynting theorem for the local energy in the sharp-time limit $s=0$, confirming that \it active energy can be localized.\rm\footnote{No such connection exists between the time-harmonic Poynting theorem and the real Poynting theorem.
}

Can reactive energy be similarly localized? The conservation theorem \eq{Q} for reactive energy now reads
\begin{align}\lab{Q2}
\pl_s (\5X_1+\5X_2)=\tfrac12\div\{\mb E_2\times\mb H_1+\mb E_1\times\mb H_2\}+\tfrac12\mb E_2\cdot\mb J_1
-\tfrac12\mb E_1\cdot\mb J_2.
\end{align}
Although the left side splits into local and nonlocal parts \it at any fixed $s>0$, \rm the right side does not. Thus, while reactive energy can be decomposed into semi-local and nonlocal parts at any fixed scale $s$, the two parts necessarily get mixed upon any change of scale. 

Hence \it there is no way to separate the local part of the reactive energy from the nonlocal part and there is no such thing as \it instantaneous reactive energy. \rm In fact, the reactive energy flux $\mb T$ and the reactive power density $\5Q$ are \sl intrinsically nonlocal \rm since they involve the Hilbert transforms $\mb E_2, \mb H_2, \mb J_2$.

This conclusion seems to contradict some popular reactive energy and power models, for example \ci{AWA7}. The confusion seems to stem from the fact that a mere \it time-dependence \rm need not signify instantaneity. Thus, although our theory gives time-dependent expressions for reactive energy and its flow, the above analysis shows that the time variable in $\5X, \mb T, \5Q$ (even with $s=0$) is nonlocal since these expressions necessarily include Hilbert transforms. This conclusion is consistent with our intuition, which says that reactive energy cannot be strictly localized in time because it necessarily involves \it lags and leads. \rm 

\rem\rm\label{R:nonloc2}
It is generally accepted that physical theories are \it local, \rm meaning that an event at $\rt$ can have an influence at $(\3r', t')$ only if $|\3r'-\3r|\le c(t'-t)$.  So  does the reactive energy and power render our theory invalid? Not at all. While $\mb E, \mb H, \mb J$ are non-local with respect to $\3E, \3H, \3J$, they satisfy the (local) Maxwell equations \eq{M} \it amongst themselves. \rm Thus if $\3J=\30$ for $t\notin [t_1, t_2]$ but $\mb J\ne\30$ there, this is not a problem as long as we don't claim that  $\mb E, \mb H$ satisfy \eq{M} with $\3J$ as the current density, and no such claim is made. 

\rem\rm\label{R:modes}  The analytic continuation  $t\to t+is$ blurs  time. This in turn relaxes the \it instantaneous \rm energy conservation law \eq{RP} to a conservation of \it average \rm energy over a window defined by the Cauchy kernel \eq{C}. Hence the analytic continuation creates a \it banking system \rm where energy can be exchanged not only across space but also across \it time \rm within the allowed window. Such energy exchanges go undetected at the given scale, so they can be called \it virtual. \rm The analytic continuation thus results in a system of \it virtual lags and leads \rm controlled by the scale parameter $s$. This also explains why reactive energy can be negative as well as positive, a natural property in a banking system.

\section{Integral conservation laws and interpretation}\label{S:interpretation}

The first conservation law \eq{P} has the same form as Poynting's theorem \eq{RP} \ci{J99}, where $U, \3S, P$ are the local versions of $\5U,\mb S,\5P$. This is similar to the situation in the time-harmonic complex Poynting theorem \eq{CP}. But  $\5U,\mb S,\5P$ are time dependent, whereas their counterparts in \eq{CP} are \it period averages. \rm There is no `period' to average over in \eq{P}, but the Cauchy kernel $C_s$ \eq{CPH} provides a natural \it scaling window \rm of width $\5O\0s$. 

The most unusual feature of the second conservation law \eq{Q} is that the parameter in which something is to be conserved is not the \it time \rm $t$, as usual, but the \it scale \rm $s$. This seems strange mainly because we are so used to things being conserved in $t$. To better understand this, we study the integral versions of the conservation laws.

Let $V$ be a bounded volume not containing any field singularities, with boundary $\pl V$ (which may consist of several pieces). Integrating \eq{P} and \eq{Q} over $V$ and applying the divergence theorem gives
\begin{equation}\lab{ints}\begin{split}
\pl_t\int_V\dd V\,\5U(\3r, t, s)&=\int_V\dd V\,\5P(\3r, t, s)-\int_{\pl V}\dd\3A\cdot\mb S(\3r, t, s)\\
-\pl_s\int_V\dd V\,\5X(\3r, t, s)&=\int_V\dd V\,\5Q(\3r, t, s)-\int_{\pl V}\dd\3A\cdot\mb T(\3r, t, s),
\end{split}\end{equation}
where $\dd\3A$ is the outward-oriented surface element of $\pl V$. By our assumptions on $V$, all the integrals converge. With obvious notation, \eq{ints} states that
\begin{equation}\lab{cons}\begin{split}
\pl_t \rrm U(t, s)&= \rrm P(t, s)- \rrm S(t, s)\\
-\pl_s \rrm X(t, s)&= \rrm Q(t, s)- \rrm T(t, s),
\end{split}\end{equation}
where $(\rrm U,\rrm X)$ are the active and reactive energies in $V$,  $(\rrm P, \rrm Q)$ are the respective powers generated in $V$, and $(\rrm S, \rrm T)$ are the respective power losses through $\pl V$.

The first law in \eq{cons} has a traditional interpretation: the rate at which the active energy $\rrm U$ in $V$ increases with $t$ equals the rate at which it is being generated, minus the rate at which it flows out through the boundary.

The second law in \eq{cons} \it seems \rm to say that the rate at which the reactive energy $\rrm X$ in $V$ \it decreases \rm with $s$ equals the rate at which it is being generated, minus the rate at which it flows out through the boundary. This interpretation is obviously wrong. For example, if reactive energy is being created in $V$ $(\rrm Q>0)$ and is also flowing \it into \rm $V$ through the boundary ($\rrm T<0$), we expect $\pl_s \rrm X(t, s)$ to be positive, but according to \eq{cons} it is negative. Of course, we could have defined $\5X=\5W_e-\5W_m$ in \eq{SUP}, in which case there would not have been a sign difference between \eq{P} and \eq{Q}. But this goes against common wisdom since a magnetic load makes the current \it lag \rm behind the voltage, thus generating \it positive \rm reactive energy, and an electric load creates a \it lead \rm in the current over the voltage, thus generating \it negative \rm reactive energy.

To resolve this puzzle, integrate \eq{cons} over the interval $0<s<s_1$:
\begin{align}\lab{cons2}
 \rrm X(t, s)- \rrm X(t, s_1)=\int_s^{s_1}\dd s'\,\{\rrm Q(t, s')- \rrm T(t, s')\}.
\end{align}
By \eq{lims},\footnote{The factor 1/2 simply means that $\5X$ and $\rrm X$ live in the world of analytic signals and $X\dc$ lives in the `RMS world' of local fields; see Remark \ref{R:RMS}.
}
$s_1\to\infty\imp\5X\to \tfrac12X\dc,\  \5Q\to 0,\  \mb T\to\30$, so
\begin{align*}
\rrm X(t, s_1)\to\tfrac12\int_V\dd V\,X\dc\orr\=\rrm X\dc, \  \rrm Q(t,s_1)\to 0,\  \rrm T(t,s_1)\to 0,
\end{align*}
where $\rrm X\dc$ is the \it static \rm reactive energy in $V$. Letting $s_1\to\infty$ in \eq{cons2} gives
\begin{align}\lab{Q3}
 \rrm X(t, s)=\rrm X\dc+\int_s^\infty\dd s'\,\{\rrm Q(t, s')- \rrm T(t, s')\}\=\rrm X\dc+\rrm X'(t, s),
\end{align}
which has the correct sign.

The resolution of the puzzle is that the proper \it orientation \rm for the scale parameter $s$ is \it backwards, \rm from the \it coarsest \rm ($s=\infty$) to the \it finest \rm scale ($s=0$). Nothing of interest happens at $s=\infty$, where all time dependence has died out. Thus $\5Q(\3r, t, s)$ should be interpreted as the rate at which reactive energy density is being generated at $\rt$ with \it decreasing \rm $s$, or equivalently at which it is being \it drained with increasing \rm $s$. Similarly, $\div\mb T(\3r, t, s)$ is the rate at which reactive energy is springing out from $\rt$ with decreasing $s$, or  \it sinking into \rm $\rt$ with increasing $s$. If $\rrm Q>0$ and $\rrm T<0$, then the reactive energy in $V$ is \it increasing \rm with decreasing $s$, so $-\pl_s \rrm X>0$ as required by \eq{cons}.

\rem\rm\label{R:cumulative} Equation \eq{Q3} shows that $\rrm X(t, s)$ is the \it cumulative reactive energy in $V$ for all scales \rm $s'\ge s$, including the static value $\rrm X\dc$. Since the variables $\mb E, \mb H, \mb J, \uprho$ are all tied together by Maxwell's equations, they must \it all \rm be interpreted as cumulative  for all scales $s'\ge s$. This makes precise the statement \eq{dt>s} that $\mb X(\3r, t+is)$ cannot include spikes in $t$ \it narrower \rm than $\5O\0s$.

\section{The scale parameter as reactive time}\label{S:rtime}

Equation \eq{cons2} shows that  the rates $\rrm Q$ and $\rrm T$, hence also $\5Q$ and $\mb T$, are \it with respect to $s$ \rm and not $t$. The symmetry between the
conservation laws \eq{P} and \eq{Q} suggests the interpretation of $s$ as a new variable, \it reactive time, \rm whose purpose is to track lags and leads, to be measured, say, in \it seconds reactive \rm [sr].
Then the MKS units of $\5Q,\mb T$, and $\rrm Q$ are
\begin{align}\lab{units}
[\5Q(\3r, t, s)]={\rm [J/m^3/sr]},&& [\mb T(\3r, t, s)]={\rm [J/m^2/sr]},&& [\rrm Q]={\rm [J/sr.]}
\end{align}
Thus, \it reactive current \rm should be measured in \it coulombs per second reactive \rm [C/sr] or \it ampere reactive \rm [Ar]. But the conventional units of reactive power are \it volt-ampere reactive \rm [VAr], as opposed to \it Watt \rm [W]= [J/sr] for \it active \rm power.  Thus
\begin{align}\lab{var1}
\rm 1VAr=1V\cdot C/sr=1J/sr,
\end{align}
in agreement with \eq{units}. The concept of reactive time thus provides a logical basis for the \it ad hoc \rm [VAr] units. From our point of view, $\5P$ and $\5Q$ are simply the rates at which energy density is pumped into the system with respect two different time variables. 

\rem\rm\label{R:noCxenergy} The fact that there is no `complex energy density' in \eq{CPT} analogous to $\mb S+i\mb T$ and $\5P+i\5Q$ shows that \it energy is energy, \rm \ie there is  no \it qualitative \rm difference between active and reactive energy; both are measured in Joules. The sole difference is \it quantitative: \rm  $\5W_e$ contributes positively to $\5U$ and negatively to $\5X$, whereas $\5W_m$ contributes positively to both. It is only where \it rates \rm of energy flow are concerned that a qualitative difference emerges between active and reactive variables, as it becomes necessary to specify if the rates are with respect to $t$ or $s$. Since $s$ is \it imaginary \rm time, this also explains why the reactive rates enter as imaginary variables in \eq{cx1}.

\section{Some possible applications}\label{S:apps}

Reactive energy has become a  popular topic in the electric power industry over the past several decades. To see why,  note that a compact fluorescent lamp with an inductive ballast generates reactive power, whereas  an incandescent lamp is purely resistive and consumes only \it active \rm power. Most digitally controlled appliances, such as computer monitors, are also reactive, and this has created problems for the energy industry. Although reactive energy performs no `useful' work, an abundance of it can put strains on the electrical grid by distorting the sinusoidal character of the current and voltage waveforms and thus interfering with the smooth flow of energy, heating transmission lines and possibly even causing brownouts or blackouts. In the absence of a solid theory, it is not even obvious how to reliably \it measure \rm reactive energy, and this leads to problems with monitoring and billing. In large-scale industrial environments, it is especially desirable to measure reactive energy flow  in \it real time\rm\,\footnote{By now we know that reactive energy cannot \it truly \rm be measured in \it real time \rm because it necessarily involves lags and leads. The local expression $X\rt$ \eq{Xt} is identified with reactive energy only because $\5X(\3r, t,s)$ is, and even $\5X(\3r, t,0)$ is non-local. Thus `real time' here means \it scaled \rm time.  Just how `real' (local) it is depends on the frequency spectra of the fields, which determine the distribution of time lags in \eq{lag}, as well as the control parameter $s$, which suppresses higher modes.
}
so that it can be corrected, thus restoring the power quality \it in-house. \rm This process, called \it reactive power conditioning, \rm is similar to matching the impedance of a load in a circuit in order to minimize the strain on the system and maximize the real power transfer. It amounts to equalizing the magnetic (inductive) energy and the electric (capacitative) energy, thus minimizing  $|\rrm X'|$ in \eq{Q3}.

For approximately sinusoidal voltage and current waveforms, there seems to be reasonable agreement in the literature on the correct treatment of reactive energy, based on the time-harmonic complex Poynting theorem. But as noted above, modern electrical grids are subject to strong non-sinusoidal influences, and the lack of a proper time-domain theory has given rise to various attempts to solve the problem at a circuit level without recourse to Maxwell's equations. There is a variety of theories but, it seems, little agreement on which one is correct.  See \ci{AWA7,C4} for  a small sample of the literature.

To apply our theory to an electric power grid, it must be reduced from the \it field level \rm to the \it circuit level. \rm As  explained in \ci[Section 6.5]{FCA60}, this requires slowing down the system to a \it quasi-static\,\rm\footnote{In \ci{FCA60}, the `speed' of time flow is controlled by the transformation $t\to t/\a, \a>0$, and the quasi-static approximation is obtained by letting $\a\to0$.  In our case it is achieved by the analytic continuation $t\to t+is$, and the quasi-static approximation is obtained by letting $s\to\infty$. It would be interesting to compare these two methods in other situations.
}
time scale which is large compared to the maximum time delay in communications across the entire system. In our terms, that means choosing a time scale
\begin{align}\lab{slc}
s\gg \ell/c,
\end{align}
where $\ell$ is a length scale representing the size of the system. In this regime all time retardation can be ignored. Thus $t$ can be taken as a systemwide `absolute time,' and there is no radiation. Under certain additional conditions, the system can be effectively described by a circuit with `lumped'  inductors, capacitors, resistors and power sources. For example, a local region is represented by an inductor if $\5X>0$ and by a capacitor if $\5X<0$.\footnote{To obtain resistors, one must add a constitutive relation $\3E=\s\inv\3J$.
}


Reactive energy also plays an important role in \it radiating \rm systems such as antennas. Radiation occurs at higher frequencies, hence at \it finer scales. \rm  As noted in \eq{far1}, reactive energy does not radiate. Therefore the efficiency of an antenna can be improved by minimizing its reactive energy, which can be done by adding a load that cancels its reactance. Perhaps the new conservation laws can help improve the design and control of wideband and pulsed antennas. 

%

\section*{Acknowledgements}
This work was supported by AFOSR Grant \#FA9550-12-1-0122. I thank Dr. Arje Nachman for his support, and Drs. Richard Albanese, Andrea Alu, Thorkild Hansen and Arthur Yaghjian for helpful discussions.

\end{document}